\documentclass[11pt,draft]{article}
 \usepackage{amsfonts}
 \usepackage{amssymb}
 \def\bdt{\dot \beta}
 \def\adt{\dot \alpha}

 \newfont{\bbbold}{msbm10}
 \def\com{\mbox{\bbbold C}}

 \def\bbC{\mbox{\bbbold C}}

 \def\cO{{\cal O}}

 \def\cR{{\cal R}}

 \def\cV{{\cal V}}

 \newfont{\goth}{eufm10 scaled \magstep1}

 \def\gg{\mbox{\goth g}}

 \def\gl{\mbox{\goth l}}

 \def\gp{\mbox{\goth p}}

 \def\gs{\mbox{\goth s}}

 \def\a{\alpha}
 \def\b{\beta}
 \def\c{\gamma}
 \def\d{\delta}\def\D{\Delta}
 \def\e{\epsilon}

 \def\l{\lambda}
 \def\m{\mu}
 
 \def\p{\pi}

 \def\t{\tau}

 \def\be{\begin{equation}}\def\ee{\end{equation}}
 \def\bea{\begin{eqnarray}}\def\eea{\end{eqnarray}}
 \def\ba{\begin{array}}\def\ea{\end{array}}

 \def\del{\partial}

 \def\unB{\underline B}

 \def\str{\rm str}
 \def\xz{\times}

 \def\del{\partial}
 
 \def\bt{\bullet}
 \def\3dt{\dot{3}}
 \def\hy{\hat{y}}

 \let\la=\label

 \let\bm=\bibitem{}

 \def\nn{\nonumber}
 \def\bd{\begin{document}}
 \def\ed{\end{document}}
 \def\bea{\begin{eqnarray}}\def\barr{\begin{array}}\def\earr{\end{array}}
 \def\eea{\end{eqnarray}}
 \def\ft#1#2{{\textstyle{{\scriptstyle #1}\over {\scriptstyle #2}}}}
 \def\fft#1#2{{#1 \over #2}}
 \newcommand{\eq}[1]{(\ref{#1})}
 \def\eqs#1#2{(\ref{#1}-\ref{#2})}
 \def\det{{\rm det\,}}
 \def\tr{{\rm tr}}\def\Tr{{\rm Tr}}

\begin{document}

 \thispagestyle{empty}

 \hfill{KCL-TH-01-??}

  \hfill{\today}

 \vspace{20pt}

 \begin{center}
 {\Large{\bf OPEs and 3-point correlators of protected operators\linebreak
\vskip .05cm  in $N=4$ SYM}}
 \vspace{30pt}

 {P.J. Heslop and P.S. Howe} \vskip 1cm {Department of Mathematics}
 \vskip 1cm {King's College, London} \vspace{15pt}

 \vspace{60pt}

 {\bf Abstract}

 \end{center}

Two- and three-point correlation functions of arbitrary protected
operators are constructed in N=4 SYM using analytic superspace
methods. The OPEs of two chiral primary multiplets are given. It is
shown that the $n$-point functions of protected operators for
$n\leq4$ are invariant under $U(1)_Y$ and it is argued that this
implies that the two- and three-point functions are not
renormalised. It is shown explicitly how unprotected operators can
be accommodated in the analytic superspace formalism in a way which
is fully compatible with analyticity. 
Some new extremal correlators are exhibited.

 {\vfill\leftline{}\vfill \vskip  10pt

 \baselineskip=15pt \pagebreak \setcounter{page}{1}

\section{Introduction}
There has been a resurgence of interest in four-dimensional
superconformal field theories over the past few years largely due
to the impact of the Maldacena conjecture \cite{maldacena} and this
has led to the discovery of many new and interesting results. Most
of these results have concerned properties of short (series C)
operators and their correlation functions derived both directly in
field theory and from supergravity via the AdS/CFT correspondence.
Some recent reviews and lists of references can be found in
\cite{agmo,f,b,hw}. These operators are protected from
renormalisation in the sense that their shortness implies that they
cannot develop anomalous dimensions because the representations
under which they transform determine these dimensions uniquely.
Recently it has become apparent from various calculations that
certain series A operators also have vanishing anomalous dimensions
\cite{afp1,afp2,epss,aeps,bkrs0,aes}. In a recent note \cite{hesh1}
the current authors have argued that the reason for this is
fundamentally the same as for the series B and C operators - the
operators concerned are short and this shortness is preserved in
the interacting quantum theory provided that the constraints
satisfied by the superfields (in super Minkowski space) are
respected by gauge invariance (see also \cite{es} and comments on
this result in \cite{ps}). One way of saying this is to note that
all composite operators can be realised as superfields on analytic
superspace and the protected ones are those that are realised as
analytic tensor superfields in the interacting quantum theory. The
non-protected operators, which do acquire anomalous dimensions,
have super Dynkin labels which include positive real numbers and it
is this fact which stops them from being realisable as normal
tensor fields.

The idea of using harmonic superspace methods \cite{gikos} to study
four-dimensional superconformal field theories was advocated in a
series of papers \cite{hw2,hsw,ehssw} and was motivated by the
realisation that the on-shell N=4 super Yang-Mills field strength
superfield can be described as a (covariantly) analytic field on a
certain harmonic superspace \cite{hh} (see also \cite{bandos}). It
was realised that there is a family of gauge-invariant scalar
multiplets which can be written as analytic superfields and which
therefore seemed to be a natural set of objects to study in field
theory. It was later observed that this set of multiplets is
actually in one-to-one correspondence with the Kaluza-Klein
multiplets of IIB supergravity on $AdS_5\xz S^5$ \cite{af2}. It has
subsequently been shown that various other short multiplets in the
theory can be written as harmonic superfields on various harmonic
superspaces \cite{afsz,fs,hesh2}. However, until recently, not much
attention has been paid to analytic superfields which carry
superindices. It is the case, however, that all representations in
four-dimensional SCFT with extended supersymmetry are carried by
analytic superfields of various types \cite{hesh3}. Moreover, in
\cite{hw} it was noted that the Konishi multiplet in the free
theory can be realised as an analytic tensor superfield.

In this paper we take the study of analytic operators a step
further by constructing the 3-point functions of analytic tensor
superfields and by identifying the leading contributions of
analytic tensor operators to the OPEs of two other such operators.
In the interacting theory this analysis is therefore relevant to
the protected  operators - all series B and C operators in the
theory and a subset of series A operators. There is a formal
resemblance between the N=4 analytic superspace we shall use and
Minkowski space which is due to the fact that they can both be
presented as subsets of Grassmannians and which allows us to adapt
the techniques of \cite{op} in a reasonably straightforward manner.
Using these methods we derive the complete expressions for the
three-point functions and the related OPEs. We also show that the
correlation functions of protected operators for $n\leq 4$ points
are invariant under $U(1)_Y$ \cite{i,is} and argue that the two- and
three-point functions of such operators are not renormalised. In
addition, we discuss how unprotected operators can be accommodated
in the analytic superspace formalism by means of an $N=2$ example.
It turns out that such operators, which we call quasi-tensors, are
perfectly compatible with analyticity in the internal space, but at
the same time allow the introduction of non-integral powers of
$x^2$ reflecting the occurrence of anomalous dimensions. These
results suggest that the formulae we have given for the OPEs and
three-point functions may also be valid, when interpreted appropriately, for arbitrary operators and
not just the protected ones. In a recent paper, Eden and Sokatchev
\cite{es} have studied some of these three-point functions and OPEs
in harmonic superspaces but in a somewhat different approach to
that adopted here. In the following we confirm some of their
results, obtained by studying the constraints that analyticity
imposes on three-point functions, in our formalism. These results
can also be obtained directly from the OPE. We also argue that there are extremal correlators involving protected operators other than chiral primaries \footnote{We use the term chiral primary to mean a supermultiplet whose leading component is a trace of $p$ factors of the gauge multiplet scalars in the representation $[0p0]$ of $SU(4)$. There are other multi-trace multiplets which transform under the same representation of the superconformal group and which have the same properties as far as the results of this paper are concerned.}
and discuss an example following the ideas of \cite{es}.

\section{Composite operators}

We begin by recalling a few facts about composite operators in
four-dimensional SCFT. The representations of the superconformal
group are well-known \cite{screp} and their realisations on
superfields have been studied by many authors, see for example
\cite{af1,fz,hh,afsz,fs,hesh2}. The quantum numbers specifying a
representation of the $N=4$ superconformal group are
$(L,J_1,J_2,a_1,a_2,a_3)$ where $L$ is the dilation weight, $J_1$
and $J_2$ are spin labels and $(a_1,a_2,a_3)$ are $SU(4)$ Dynkin
labels. The unitarity bounds for the three series of operators are

 \be
 \ba{lll}
 {\rm Series\ A:} & L\geq 2+2J_1 + 2m_1 -{m\over2}\qquad &L\geq
 2+2J_2 +{m\over2} \\
 &&\\
 {\rm Series\ B:} & L= {m\over2};\ & L\geq 1+m_1 + J_1,\  J_2=0\ \
 {\rm
 or}\\
 &&\\
 &L=2m_1-{m\over2}; & L\geq 1 + m_1 + J_2,\ J_1=0\\
 &&\\
 {\rm Series\ C:} & L=m_1={m\over2} &J_1=J_2=0
 \ea
 \la{1}
 \ee

where $m$ is the total number of boxes in the Young tableau of the
$SU(4)$ representation and $m_1$ the number of boxes in the first
row.

It will be useful later on to be able to write these
representations in terms of super Dynkin diagrams. For the
(complexified) superconformal group $SL(4|N)$ acting on
$\com^{4|N}$, the Dynkin diagram depends on the choice of basis. If
the basis is ordered in the standard fashion, 4 even - N odd, we
have the distinguished basis with one odd root, but we shall use a
different basis, which we shall refer to as physical, in which the
basis has the ordering, 2 even - N odd - 2 even. The physical basis
has two odd roots so that the Dynkin diagram is
\be
\begin{picture}(220,20)(-10,-10)
\put(0,0){\makebox[0pt][l]{$\bt\hspace{1.5em}\ominus\hspace{1.5em}
    \underbrace{\bt \hspace{1.5em}
\bt\hspace{1.5em}\cdots
\hspace{1.5em}\bt\hspace{1.5em}\bt}_{N-1}\hspace{1.5em}\ominus
\hspace{1.5em} \bt$} \rule[.5ex]{6.85em}{.1ex} $\hspace{4.5em}$
\rule[.5ex]{6.6em}{.1ex}
 }
\end {picture}
\la{2} \ee

Any representation can be specified by giving labels associated to
each node of the Dynkin diagram. The labels associated with the two
external even (black) nodes are determined by the spin quantum
numbers $(J_1,J_2)$ and the $(N-1)$ internal even labels are fixed
by the Dynkin labels of $SL(N)$. The two odd (white) labels are
then determined by the dilation ($L$) and the R-symmetry ($R$)
quantum numbers. All the Dynkin labels should be non-negative
integers except for the odd ones which can be positive real
numbers. These continuous labels are directly related to anomalous
dimensions of operators.

In order to have unitary representations (of the real
superconformal group $SU(2,2|N)$) the Dynkin labels on the odd
nodes must exceed those of the adjacent external nodes by at least
one unless one or both pairs of these adjacent nodes are zero. This
gives three series of unitary bounds. We label the nodes from the
left $n_1 \dots n_{N+3}$ so that the two odd nodes are $n_2$ and
$n_{N+2}$ and the adjacent external nodes are $n_1$ and $n_{N+3}$
respectively. For series A we have $n_2 \geq n_1 +1$ and $n_{N+3}
\geq n_{N+2}+1$. For series B we have either $n_1=n_2=0$ and
$n_{N+3} \geq n_{N+2}+1$ or we have $n_2 \geq n_1 +1$ and $n_{N+3}
= n_{N+2}=0$. Finally series C requires that $n_1=n_2=n_{N+3} =
n_{N+2}=0$. For general $N$ we have

\bea n_2&=&{1\over2}(L-R) + J_1 + {m\over N} - m_1 \nn\\
n_{N+2}&=&{1\over2}(L+R) + J_2 -{m\over N} \la{2.1} \eea

where $m$ is the total number of boxes in the internal Young
tableau determined by the $SU(N)$ Dynkin labels $(a_1,\ldots
a_{N-1})=(n_3,\ldots n_{N+1})$ and $m_1$ is the number of boxes in
the first row. The external black labels are
$(n_1,n_{N+3})=(2J_1,2J_2)$. For $N=4$ we need to impose $R=0$ in
order to have representations of $PSU(2,2|4)$. This implies that

 \be
 n_3+ 2n_2 -n_1=n_5 + 2n_6 -n_7
 \la{3}
 \ee

The super Dynkin diagram can also be used to represent coset spaces
determined by parabolic subgroups. With respect to a given basis
the Borel subalgebra consists of lower triangular matrices, and a
parabolic subalgebra (which by definition is one which contains the
Borel subalgebra) consists of lower block triangular matrices. The
size of these blocks is determined by a set of at most $N+3$
positive integers $k_1 < k_2 \dots$ and can be represented on the
Dynkin diagram by placing crosses through the $k_i$th nodes
(starting from the left). For example, super Minkowski space is
represented by
\be
\begin{picture}(220,0)(-5,0)
\put(0,0){\makebox[0pt][l]{$\bt\hspace{1.5em}\otimes\hspace{1.5em}
\bt\hspace{1.5em} \bt\hspace{1.5em}\cdots
\hspace{1.5em}\bt\hspace{1.5em}\bt\hspace{1.5em}\otimes
\hspace{1.5em} \bt$} \rule[.5ex]{6.7em}{.1ex} $\hspace{4.5em}$
\rule[.5ex]{6.9em}{.1ex}
 }
\end {picture}
\la{4} \ee

Chiral superspaces have a single cross through one of the odd
nodes, harmonic superspaces have crosses through both odd nodes and
some internal nodes, and analytic superspaces have crosses only
through internal nodes. The superspace of most interest to us in
this paper is $(N,p,q)=(4,2,2)$ analytic superspace; the Dynkin
diagram is:
\be
 \begin{picture}(192,10)(0,0) \put(0,0){\makebox[0pt][l]{$\bt
 \hspace {2em}
 \ominus \hspace{2em}\bt\hspace{2em}\times\hspace{2em}\bt
 \hspace{2em}\ominus \hspace{2em}\bt$}
 \rule[.5ex]{17.1em}{.1ex} }
 \end{picture}
 \la{5}
 \ee

 This space is a subset of a super Grassmannian with local coordinates

 \be
 X^{AA'}=\left( \ba{cc} x^{\a \adt} & \l^{\a a'} \\
                        \pi^{a\adt}  &  y^{a a'}  \ea
 \right).
 \la{6}
 \ee

where $x$ denotes the spacetime coordinates, $\l,\pi$ are odd
coordinates and $y$ are coordinates for the internal manifold. The
indices $(\a,\adt)$ are 2-component spacetime spinor indices while
$(a,a')$ are 2-component spinor indices for the internal space
which is (locally) the same as spacetime in the complexified case.
The capital indices span both spacetime and internal indices,
$A=(\a,a),\ A'=(\adt,a')$, and we use the convention that
$(\a,\adt)$ are even indices while $(a,a')$ are odd. An important
feature of analytic superspace is that superfields carrying
irreducible representations are completely specified by the super
Dynkin labels and analyticity; no further constraints need to be
imposed.

The crosses on a super Dynkin diagram factorise the diagram into
sub-(super)-Dynkin diagrams corresponding to the semi-simple
subalgebra of the Levi subalgebra (the diagonal blocks in the
parabolic), while the Dynkin labels above the crosses correspond to
charges under internal $U(1)$'s or dilation and R weights. In
general the Levi subalgebra will be a superalgebra and so the
fields can carry superindices. Only in cases where both odd nodes
have crosses through (such as for super Minkowski space and
harmonic superspaces) does the Levi subalgebra contain no
superalgebra.

 In the free theory the Maxwell field strength superfield,
 corresponding to the representation with $n_4=1$ and all other Dynkin
 labels zero, is a single component analytic superfield $W$. In the
 interacting case $W$ is covariantly analytic and so is not a
 superfield on analytic superspace. However, gauge-invariant
 products of $W$ are. The operators
 $A_p:=\tr(W^p)\ p=2,3,\ldots$ which transform under the representations
 which have only the central Dynkin label
 non-zero are
 in one-to-one correspondence with the Kaluza-Klein supermultiplets  of
 IIB supergravity on $AdS_5 \xz S^5$ \cite{gun,af2,hw2}. The operator
 $A_2:=T$
 is special;
 it is the supercurrent multiplet. $A_p$ is a scalar under
 $\gs\gl(2|2)\oplus\gs\gl(2|2)$ and has charge $p$ under the $U(1)$
 corresponding to the central node of the super Dynkin diagram. All other
 representations transform non-trivially
 under the sub-algebra $\gs \gl(2|2) \oplus \gs \gl(2|2)$. The series B
 superfields must transform under the totally (generalised) antisymmetric
 tensor representation (or the trivial representation) of one of the
 $\gs \gl(2|2)$ subalgebras and the series C superfields must transform
 under the totally antisymmetric representation of both $\gs \gl(2|2)$
 subalgebras (trivially in the KK case). For a general representation
 the highest weight state is obtained from the tensor component
 which has the most number of internal ($a$ or $a'$) indices. The
 short representations are the series B and C operators and the
 series A operators which saturate at least one of the unitarity
 bounds and which can be written in terms of derivatives
 $\del_{A'A}$ acting on products of $A_p$s on analytic superspace.

Let us now consider an arbitrary analytic tensor operator $\cO$
specified by a set of super Dynkin labels. The 3 leftmost labels
determine the representation of the left $\gs\gl(2|2)$ under which
it transforms and similarly for the 3 rightmost labels. The left
(right) sets of 3 labels are associated with tensors which carry
$A(A')$ indices. The 3 left super Dynkin labels $[n_1 n_2 n_3]$
correspond to a Young tableau of the form $<n_3+n_2 -n_1,
n_2-n_1,n_1>$, where the first entry gives the number of boxes in
the leftmost column, the second the number of boxes in the second
column and third gives the number of additional columns all of
which have only one box.\footnote{This is a different convention to
that used in \cite{hesh1}.} In a similar manner the 3 right super
Dynkin labels $[n_5,n_6,n_7]$ correspond to the Young tableau
$<n_5+n_6-n_7, n_6-n_7,n_7>$. We shall denote these representations
by $\cR$ and $\cR'$ respectively. The total number of indices of
representation $\cR$ is the same as the number of boxes in the
Young tableau and is given by $S:=n_3 + 2n_2-n_1$. This number is
the same for both $\cR$ and $\cR'$ because we are concerned only
with representations for which $R=0$. These have the same number of
primed and unprimed indices.

We shall denote a general operator by $\cO_{\cR \cR'}^Q$ where
$Q=L-(J_1+J_2)$. The transformation rule for such an operator under
an infinitesimal superconformal transformation is

 \be
 \d \cO_{\cR \cR'}^Q = \cV \cO_{\cR \cR'}^Q + \cR(A(X)) \cO_{\cR
 \cR'}^Q + \cR'(D(X)) \cO_{\cR \cR'}^Q + Q\D \cO_{\cR \cR'}^Q
 \la{7}
 \ee

where $\cV$ is the vector field generating the transformation,
$A(X)$ and $D(X)$ are $X$-dependent parameters for the left and
right $\gg\gl(2|2)$ algebras, $\cR(A(X))$ and $\cR'(D(X))$ extend
the representations $\cR,\cR'$ to $\gg\gl(2|2)$ in a natural way,
and $\D=\str(A + XC)$. Here,

 \bea
 \cV X &=& B + AX + XD + XCX\\
 A(X)&=&A + XC \\
 D(X)&=& D + CX
 \la{8}
 \eea

and $A,B,C,D$ are super-matrices parametrising an infinitesimal
$\gs\gl(4|4)$ matrix $\d g$,

 \be
 \d g=\left(
 \ba{rr}
 -A^A{}_B & B^{A B'}\\
 -C_{A' B} & D_{A'}{}^{B'}
 \ea\right)
 \la{9}
 \ee

Note that this defines a transformation of $\gp\gs\gl(4|4)$ on
analytic superspace because the element proportional to the unit
matrix does not act.

\section{Two-point functions}

The conjugate representation is given by reversing the super Dynkin
labels of a given representation, and to get non-vanishing
two-point functions we have to pair an operator and its conjugate.
Thus we have

 \be
 <\cO^Q_{\cR\cR'}(1)\cO^Q_{\cR'\cR}(2)>\sim (g_{12})^Q
 \cR(X_{12}^{-1}) \cR'(X_{12}^{-1})
 \la{10}
 \ee

In this expression $X_{12}:=X_1-X_2$ as usual, and the propagator
$g_{12}$ is defined by

 \be
 g_{12}:={\rm sdet} X_{12}^{-1}={\hat y_{12}^2\over
 x_{12}^2}={y_{12}^2\over \hat x_{12}^2}
 \la{11}
 \ee

where

 \bea
 \hat x_{12}&=& x_{12} - \l_{12} y_{12}^{-1} \p_{12}\\
 \hat y_{12}&=& y_{12} - \p_{12} x_{12}^{-1} \l_{12}
 \la{12}
 \eea

In these expressions matrix multiplication is understood, the
inverses having downstairs indices
$(x^{-1})_{\adt\a},(y^{-1})_{a'a}$. The meaning of the $\cR$
symbols is as follows: at point 1 the unprimed indices on the
operator accord with the representation $\cR$, so one takes $S$
factors of $X_{12}^{-1}$ (which has downstairs indices) with the
unprimed indices in that representation. The primed indices then
have to be in the same representation, and then similarly for
$\cR'$. For example, consider the series C operators. These have
zero for the external 2 nodes on either side and so are completely
specified by their $SU(4)$ Dynkin labels which have to be of the
form $[qpq]$. If we take $q=1$ for simplicity then the
corresponding operator is a covector $\cO_{AA'}$ on analytic
superspace with dilation weight $L=p+2=Q$. In this case we get

 \be
 <\cO^{p+2}_{AA'}(1)\cO^{p+2}_{BB'}(2)>\sim (g_{12})^{p+2}
 (X^{-1}_{12})_{A'B}(X^{-1}_{12})_{B' A}
 \la{13}
 \ee

The reason this formula works is because the variation of $X_{12}$
can be written

 \bea
 \d X_{12}& = &A_1 X_{12} + X_{12} D_2 \\
           & = &A_2 X_{12} + X_{12} D_1
 \la{14}
 \eea

where $A_1:=A(X_1)$, etc. Thus, applying a superconformal
transformation to the above two-point function we find that it is
invariant because the propagator factors take care of the dilations
and the ``rotation'' factors ($A(X),D(X)$) are absorbed by the
$X$s.

As the inverses of coordinate matrices $X$ appear repeatedly in the
correlation functions it is worthwhile giving the exact form,

 \be
 X^{-1}=\left(\ba{rr}
 \hat x^{-1} & -x^{-1}\l \hat y^{-1} \\
 -y^{-1} \p\hat x^{-1} & \hat y^{-1}
 \ea\right)
 \la{15}
 \ee

In order to study possible singularities in the internal
coordinates it is more useful to express $X^{-1}$ in terms of $x$
and $\hat y$, and to write the propagator in terms of the same
variables. We find

 \bea
 \hat x &=& x(1+x^{-1}\l\hat y^{-1}\p)^{-1}\\
y^{-1} \p\hat x^{-1}&=& \hat y^{-1}\p x^{-1}
 \la{16}
 \eea

so that each term in $X^{-1}$ behaves at worst like $\hat y^{-1}$,
although there are nilpotent factors in all of the singular terms
except for the $y$ entry.

\section{Three-point functions}

\subsection{The general case}

We now turn to three-point functions. We can adapt the Minkowski
spacetime formalism of \cite{op} more or less straightforwardly to
analytic superspace. The idea is to use factors of $X_{12}^{-1}$
and $X_{13}^{-1}$ to translate points $2$ and $3$ to point 1, that
is, we write

 \bea
 <\cO^{Q_1}_{\cR_1\cR'_1}(1)\cO^{Q_2}_{\cR_2\cR'_2}(2)
 \cO^{Q_3}_{\cR_3\cR'_3}(3)>&\sim&(g_{12})^{Q_{12}}(g_{23})^{Q_{23}}
 (g_{31})^{Q_{31}}\qquad \xz \\
 &\phantom{\sim}&\cR_2(X_{12}^{-1})\cR'_2(X_{12}^{-1})
 \cR_3(X_{13}^{-1})\cR'_3(X_{13}^{-1})\xz \ t
 \la{17}
 \eea

where

 \be
 Q_{ij}={1\over2}(Q_i + Q_j - Q_k),\qquad k\neq i,j
 \la{18}
 \ee

and where $t$ is a tensor that transforms under the representations
$\cR_1,\cR'_1$ and the contragredient representations of
$R_2,R_2',R_3,R_3'$, that is, the representations with the same
symmetry properties but with the indices upstairs. However, due to
the factors of $X_{12}^{-1}$ and $X_{13}^{-1}$ that have been
introduced, the tensor $t$ transforms under rotations only at point
1. In addition, the propagator factors take care of the dilation
weights. We shall obtain a solution to the Ward identities if $t$
is a tensorial function (of the desired type) of
$X_{12}^{-1}-X_{13}^{-1}$ as this combination has the property that
it transforms only by rotations at point 1. This is easy to see by
noting that it can be rewritten as $-X_{123}^{-1}$ where

 \be
 X_{123}=X_{12}X_{23}^{-1}X_{31}
 \la{19}
 \ee

Therefore, the tensor $t$ is a monomial in $X_{123}$ and
$X_{123}^{-1}$ with its indices arranged to fall into the correct
representations as described above. Roughly speaking, $t\sim
(X_{123})^{2S_2 + 2S_3} (X_{123}^{-1})^{2S_1}$. We note that there
will in general be more that one solution for a given choice of
operators, although in some cases there may be no solutions if the
representations do not match properly. In addition it may be
possible to have contractions  between upper and lower indices on
$t$ leading to smaller powers of both $X_{123}$ and its inverse.
Finally, it may be the case that there are solutions but that these
solutions exhibit singularities in the internal $y$ variables. In
such cases the coefficients must be zero.

\subsection{Chiral primaries}

In order to clarify the above we shall give a few examples.
Firstly, consider three chiral primaries $<A_{p_1}A_{p_2}A_{p_3}>$.
In this case the formula reduces to the one given before in
\cite{hw2},

 \be
 <A_{p_1}A_{p_2}A_{p_3}>\sim(g_{12})^{p_{12}} (g_{13})^{p_{13}}
 (g_{23})^{p_{13}}
 \la{19.1}
 \ee

where

 \be
 p_{ij}={1\over2}(p_i + p_j - p_k),\qquad k\neq i,j
 \la{19.2}
 \ee

Analyticity implies that the sum of the charges must be even and
that the sum of any two of the charges must be greater or equal to
the third otherwise there will be singularities in the $y$'s.

Next consider the three-point function of two chiral primaries
$A_p,A_q$ and one arbitrary operator $\cO$ (at point 1). From the
above general formula it can be seen that the factors of
$X_{12}^{-1}$ and $X_{13}^{-1}$ are not required, and so we simply
get a product of $X_{123}^{-1}$'s. In order for this to be non-zero
the operator $\cO$ must be self-conjugate. We have

\be
<\cO^Q_{\cR\cR'}(1) A_p(2) A_q(3)>\sim \cR(X_{123}^{-1})
(g_{12})^{{1\over2}(Q+p-q)} (g_{13})^{{1\over2}(Q+q-p)}
(g_{23})^{{1\over2}(p+q-Q)} \la{20} \ee

where we have used the facts that the representations $\cR$ and
$\cR'$ are the same and that the $Q$ charge of $A_p$ is $p$. Now we
need to analyse which operators are allowed by analyticity. We
recall that the representation $\cR$ can be represented by the
Young tableau $<b+c,b,a>$ where $a=n_1, b=n_2-n_1, c=n_3$. (For
self-conjugate representations we take the super Dynkin labels to
be $[a(a+b)cdc(a+b)a]$.)  Consider the expression $\cR(X^{-1})$;
its inverse is $\cR(X)$. Now the highest power of $y$ in $\cR(X)$
resides in the component  which has the maximum number of internal
indices. This is achieved by filling up the first two columns with
internal indices (recall that they are odd, so graded
symmetrisation implies actual antisymmetrisation for them). This
gives us $(y)^{2b+c}$, but actually, again because of antisymmetry,
we get $b$ contractions, so we find a term of the form $(y^2)^b
y^c$, the indices on the final $c$ factors being totally symmetric
(both primed and unprimed). There are also nilpotent terms with the
same index structure but they have the effect of amending $y$ to
$\hy$. The leading singularity for the inverse is then given by
$(\hy^2)^{-(b+c)}$. Since $X_{123}=X_{12} X_{23}^{-1} X_{31}$ we
find a leading singularity structure of $\cR(X^{-1}_{123})$ of this
type for both the $(12)$ and $(13)$ channels. This implies

 \bea
 Q +p-q- 2(b+c)&\geq &0 \\
 Q +q-p-2(b+c) &\geq& 0
 \la{21} \eea

If we assume that $p\geq q$ we therefore have

 \be
 Q\geq 2(b+c) + p-q
 \la{21.1}
 \ee

In the $(23)$ channel, on the other hand, the $\cR(X_{123}^{-1})$
factor is regular and actually softens the singularity due to the
propagators. This means that we need to look for the factor with
the smallest number of $y$'s, and this is given by the term with
the largest number of spacetime indices. The latter is $a + 4$,
provided that $b\geq2$. The number of $y$'s is $2(b-2) + c$ and the
lowest term in $y_{23}$ is therefore $(y_{23}^2)^{(b-2)}
(y_{23})^c$. There are also nilpotent contributions with the same
index structure which again have the effect of changing the $y$s to
$\hy$s.  The factor $(y_{23})^c$, which is symmetric in its $c$
primed and unprimed indices, does not soften singularities of the
form $(y_{23}^2)^{-1}$ and so we obtain the analyticity bound

 \be
 Q\leq p+q-4 +2b,\qquad {\rm for}\ b\geq 2
 \la{21.2}
 \ee

If $b=0,1$, on the other hand, then the $\cR$ factor does not
affect the $(23)$ singularities at all and so we simply get

 \be
 Q\leq p+q, \qquad {\rm for}\ b=0,1
 \la{21.3}
 \ee

The analyticity bounds are therefore \eq{21.1} together with
\eq{21.2} for $b\geq 2$ or \eq{21.3} for $b=0,1$. If $b=0$ then
necessarily $a=0$ and so the operator is series C and protected.
For such operators we have $Q=2c+d+2$ and the bounds imply that
$d\geq p-q$ while $Q=p+q-2k$ with $k=0,1,\ldots q$ and $k+c\leq q$.

If $b=1$ we again have \eq{21.3}, but now $Q=2c+d+2$, and now the
possible values of $Q$ are $Q=p+q-2k$ for $k=0,1,\ldots (q-1)$ with
$k + c\leq (q-1)$ and $d\geq p-q$. These operators correspond to
saturated series A representations, and may or may not be protected
in the interacting theory.

If $b\geq 2$ one has the bound \eq{21.2} with $Q=2b+2c+d$, $c\leq
q-2$ and $d\geq p-q$. These operators are unsaturated and are thus
expected to acquire anomalous dimensions in the interacting case.

These results are in accord with those of \cite{es} but our
interpretation of them is slightly different. The cases $b=0,b\geq
2$ are not problematic; for $b=0$ the operator at point 1 is series
C and protected, while for $b\geq 2$ the operator is unsaturated
and can acquire an anomalous dimension in the interacting theory.
In fact, for fixed values of $a$ and $c$ all of these $b\geq 2$
representations have the same number of components. It therefore
makes sense to allow $b$ to be non-integral. As we shall see in
section 6 this can be done explicitly in analytic superspace and we
can also consider values of $b$ less than $2$ down to $b=1$ at
which point these representations become reducible. This point is
relevant to the question of whether a $b=1$ operator is protected
or whether it acquires an anomalous dimension. For the latter to
occur, there must be an available representation from $b\geq2$
category which can be continuously deformed into the $b=1$
representation under consideration.

Using the bounds on the representations that can arise it is easy
to show that the only $b=1$ operators that are guaranteed to be
protected are those with the maximum possible value of $Q$, i.e.
$Q=p+q$. Since the maximum value of $Q=p+q-4+2b$ for $b\geq 2$, we
see that when we continue this formula to $b=1$ we arrive at
$Q=p+q-2$. So the $b=1, Q=p+q$ representations must be protected;
there is no allowed representation with $b=1$ replaced by $b=1+\c$.
Note also that these operators have the right quantum numbers to be
constructible explicitly from analytic superspace derivatives
acting on a product of $A_p$ and $A_q$, so that the constraints
they would obey as Minkowski space superfields are consistent with
interactions.

If we now consider sub-maximal values of $Q$ such as $Q=p+q-2$, for
$b=1$, then we see that there are representations from the $b\geq2$
series which coincide with them when $b$ is continued down to
$b=1$. Furthermore, it is straightforward to check that the $SU(4)$
quantum numbers fall into the same representations. It therefore
seems to us that it is not possible to say whether a given operator
with these quantum numbers is protected or not purely on the basis
of representations and analyticity; one needs to know whether the
operator concerned can be written as an analytic tensor in the
interacting theory or not.

To illustrate this point let us consider $p=4,q=2$. For the $b=1$
operators with $Q=p+q-2=4$, the only possible values of the $SU(4)$
Dynkin labels are $c=0,d=2$. For $b\geq 2$, we again find $c=0,d=2$
and $Q=2+2b$. Continuing this down to $b=1$ we find that $Q=4$ and
so the two representations match. If we take $a=0$ then the super
Dynkin labels of this operator are $[0b020b0]$. For $b=1$ there are
three possibilities in the free theory: $T^2$, $K\xz T$, where $K$
is the Konishi multiplet, and $W^4$, all with leading components
which are scalars in the $20'$ representation of $SU(4)$. However,
in the interacting theory only the first of these is protected as
the other two become long and indeed acquire anomalous dimensions.
In the context of the three-point functions involving $A_4$ and
$A_2(=T)$, however, any of the three is allowed to appear.

\subsection{Series C operators}

Now let us consider three-point functions of series C operators,
that is operators which have super Dynkin labels $[00cdc00]$. These
include both the single-trace KK multiplets (CPOs) and multi-trace
multiplets which can be either in the same $SU(4)$ representations
as the CPOs (i.e. $[0d0]$) or more general ones of the form $[cdc],
c\neq0$. These more general superfields can be represented as
analytic superfields on $(4,1,1)$ superspace, but on $(4,2,2)$
superspace they become analytic tensors with superindices. Such
operators have the form $O^Q_{A_1 \ldots A_c A'_1 \ldots A'_c}$
with $c$ antisymmetric unprimed superindices and $c$ antisymmetric
primed superindices; they have $Q$-charge equal to $2c+d$ which is
equal to the number of boxes, $m_1$, in the first row of the
$SU(4)$ Dynkin diagram $[cdc]$.

The general formula for series C three-point functions is

 \be \ba{l}
 \langle \vspace{2mm} O^{Q_1}_{A_1 \ldots A_{c_1}A'_1 \ldots A'_{c_1} }
 O^{Q_2}_{B_1 \ldots B_{c_2}B'_1 \ldots B'_{c_2} }
 O^{Q_1}_{C_1 \ldots C_{c_3}C'_1 \ldots C'_{c_3} }\rangle \\
 \vspace{2mm}
 =(g_{12})^{Q_{12}}(g_{13})^{Q_{13}}(g_{23})^{Q_{23}}\ \ \ \xz \\
 \vspace{2mm}
 \phantom{=}(X^{-1}_{12})_{A'B}^{a_{12}}
 (X^{-1}_{12})_{B'A}^{a_{21}}(X^{-1}_{13})_{A'C}^{a_{13}}
 (X^{-1}_{13})_{A'C}^{a_{31}}
 (X^{-1}_{23})_{B'C}^{a_{23}}(X^{-1}_{23})_{C'B}^{a_{32}}\ \ \ \xz
 \\\vspace{2mm}
 \phantom{=}(X^{-1}_{123})_{A'A}^{a_{11}}
 (X^{-1}_{231})_{B'B}^{a_{22}}(X^{-1}_{312})_{C'C}^{a_{33}}\ea
 \la{23}
 \ee

 where $(X^{-1})^{a_{12}}_{A'B}:=X^{-1}_{A'_1 B_1}\ldots
 X^{-1}_{A'_{a_{12}}B_{a_{12}}}$ with the indices antisymmetrised,
 $X_{ijk}:=X_{ij}X^{-1}_jk X_{ki}$
 and
 $\{a_{ij}\}$ is a set of nine non-negative integers such that

 \be
 \sum_{j=1}^{3} a_{ij} = c_i, \qquad \sum_{i=1}^{3} a_{ij} = c_j
 \la{24}
 \ee

 The simplest solution is $a_{ii}=c_i$, with all other $a_{ij}$
 vanishing, which corresponds to the three-point function

 \be
 \langle O_1 O_2 O_3 \rangle =
 (g_{12})^{Q_{12}}(g_{13})^{Q_{13}}(g_{23})^{Q_{23}}
 (X^{-1}_{123})_{A'A}^{c_1}
 (X^{-1}_{231})_{B'B}^{c_2}(X^{-1}_{312})_{C'C}^{c_3}
 \la{25}
 \ee

 In fact the general solution is given in terms of $a_{12}, a_{13}$
 and $a_{21}$ by

 \be
 \left( \ba{ccc} a_{11}& a_{12} & a_{13}\\
           a_{21}& a_{22} & a_{23}\\
           a_{31}& a_{32} & a_{33} \ea \right) =
 \left( \ba{ccc} c_1-a_{12}-a_{13}& a_{12} & a_{13}\\
           a_{12}& c_2-a_{12}-a_{23} & a_{23}\\
           a_{13}& a_{23} & c_3-a_{13}-a_{23}\\ \ea \right).
 \la{26}
 \ee

 with

 \be
 \ba{rcl}
 a_{12} + a_{13} &\leq& c_1\\  a_{12} + a_{23} &\leq& c_2\\
 a_{13} + a_{23} &\leq& c_3. \ea
 \la{27}
 \ee

 We now turn to the question of analyticity. In order to not have any
 poles in the internal coordinates $y_{12}$ we require that for each
 term involving $X_{12}^{-1}$ or $X_{123}^{-1}=X^{-1}_{13}-X^{-1}_{12}$
 or
 $X^{-1}_{231}=X^{-1}_{21}-X^{-1}_{23}$ there is a
 corresponding power of $(g_{12})$ to cancel the poles, and similarly
 for the
 coordinates $y_{13}$ and $y_{23}$. In other words we need

 \be
 Q_{ij} \geq a_{ij} + a_{ji}+ a_{ii} + a_{jj}
 \la{28}
 \ee

 but it is easy to see that $Q_{ij}= {1 \over 2} (d_i + d_j -d_k) +
 a_{ij} + a_{ji} + a_{ii} + a_{jj} - a_{kk}$ and thus analyticity
 gives the three relations

 \be
 d_i + d_j -d_k \geq 2 a _{kk}.
 \la{29}
 \ee

\section{OPEs}

There have been several studies of the OPE in $N=4$ SYM, see for
example \cite{ope,bkrs1,is}. In \cite{hw3} a discussion of the OPE of
two supercurrents in analytic superspace was given, but analytic
tensor superfields were not taken into account and this vitiates
somewhat the conclusions of that article. However, it has been
observed that the Konishi multiplet can be accommodated in the OPE
as an analytic tensor superfield, at least in the free theory
\cite{hw}. In a more recent paper, Eden and Sokatchev \cite{es}
have analysed which operators can be expected to appear in the OPE
of two $T$s on the basis of studying the constraints that
analyticity imposes on three-point functions with two $T$s and an
arbitrary third operator. Here we give the complete OPE of two
$T$'s, at least as far as protected operators are concerned (and
ignoring descendants). Moreover, we show that analyticity of the
OPE itself is sufficient to derive the constraints on the operators
that can appear. This is also true for general CPOs (and presumably
for other operators).

The OPE of two $T$'s has the following form:

 \be
 T(1)T(2)\,\sim\, \sum C_{\cR} \cR(X_{12})(g_{12})^{{1\over2}(4-Q)}
  \cO^Q_{\cR\cR'}(2) +
 \ldots
 \la{30}
 \ee

The operators allowed on the right-hand side have to be in
self-conjugate representations so that the indices on $\cR(X_{12})$
can hook up with the indices on the operator concerned. The
$C_{\cR}$'s are numerical coefficients, and the dots indicate the
contributions of descendants. The set of operators includes the
unit operator which contributes the central charge term of the OPE.

To analyse the restrictions due to analyticity let us suppose that
an operator on the RHS is in the representation specified by the
Young tableau $<b+c,b,a>$ for the unprimed $\gg\gl(2|2)$ (and
therefore also for the primed one). The leading singularity will
occur when $\cR(X_{12})$ has the fewest number of $y$'s allowed.
This will occur when as many boxes as possible of the Young tableau
are filled with $\a$ indices. We can distinguish three cases,
$b\geq2,b=1,b=0$.

For $b\geq 2$ we can fill up the first two rows of the Young
tableau with $\a$'s and so there can be $2(b-2)+c$ powers of $y$.
Explicitly, this gives $(y^2)^{b-2} y^{c}$. In fact, there are
nilpotent contributions with fewer $y$'s but these simply have the
effect of changing $y$ to $\hat y$. The term $y^c$, which is
symmetric on both primed and unprimed indices, cannot help with
singularities if $y^2$, so we obtain the inequality

 \be
 Q\leq 2b
 \la{31}
 \ee

Since $Q=2(b+c) + n_4$ this implies that $c=d=0$ and $Q=2b$. These
operators are unsaturated series A and have trivial internal
labels.

For $b=1$ the lowest power of $y$ we can obtain is $y^c$ and does
not affect the singularity. So in this case we find $Q\leq 4$.
Since all the super Dynkin labels are integers in this case, we can
have only $Q=4,2$. If $Q=4$ we can have either $d=2,c=0$ or $d=0,
c=1$. If $Q=2$ we must have $c=d=0$.

For $b=0$ we have $n_1=n_2=0$ so that the operator is series C.
Again we find that $Q\leq 4$ so we can have $Q=0,2,4$, and we also
have $Q=2c + d$. So for $Q=4$ we can have $d=4,c=0$ or $d=2,c=1$ or
$d=0, c=2$. For $Q=2$ we can have $d=2, c=0$ or $d=0,c=1$. Finally,
if $Q=0$ we must have $d=c=0$; this is just the unit operator.

These results are in agreement with those we obtained previously
from the three-point functions and also with those of \cite{es}. We
now comment on the interpretation. Firstly, for $b=0, Q=2$, if the
internal Dynkin labels are $[020]$ then this operator is $T$. If
the labels are $[101]$ we get an operator which doesn't exist in
the interacting theory. For $b=0,Q=4$ the operator with $SU(4)$
labels $[040]$ is $T^2$, the operator $[202]$ can be constructed
from two $T$s and two derivatives, while the operator $[121]$ would
require one derivative and two $T$s and so does not exist as a
primary field in the interacting theory.

For $b=1,c=0, d=2$ we have the family of operators with super
Dynkin labels $[a(a+1)020(a+1)a]$. These are protected operators
which can be constructed from two $T$s and $a+2$ derivatives
symmetrised with respect to both sets of indices. The lowest member
of this family, with $a=0$, is the operator whose leading component
is $T^2$ in the $20'$ representation of $SU(4)$. If $b=1, c=1, d=0$
we get another family of operators with super Dynkin labels
$[a(a+1)101(a+1)a]$. These can be constructed from two $T$s and
$(a+3)$ derivatives with both sets of indices in the representation
$<2,1,a>$. Note that all of these operators exist in the
interacting theory and can be constructed from derivatives and $T$s
so that the constraints they would obey as superfields on Minkowski
superspace are consistent with gauge invariance.

For $b=1, Q=2$ we have a family of operators with super Dynkin
labels $[a(a+1)000(a+1)a]$. In the free theory such operators can
be constructed from two $W$'s and $a+1$ derivatives. However, in
the interacting theory, $W$ carries group indices so that the
derivatives would have to be gauge-covariant. Since the gauge
potential itself is not a field on analytic superspace, it follows
that such operators cannot be constructed in the interacting theory
as analytic tensor superfields. Finally, if $b\geq 2$ we have the
family of operators with $Q=2b, c=d=0$. These have super Dynkin
labels $[a(a+b)000(a+b)a]$. Such representations make sense for
non-integral $b$ and in fact they all have the same number of
components for a given $a$. We shall argue below that it makes
sense to consider $R(X_{12})$ for these representations as long as
$b>1$. Hence, in the interacting quantum theory the $b=1, Q=2$
series operators acquire anomalous dimensions to become unsaturated
operators of the same type as the  $b\geq2, Q=2b$ operators. As an
example, consider the Konishi operator, $K=\tr(W_I W_I)$ ($I,J$
$SO(6)$ vector indices) as a superfield on Minkowski superspace. In
the free theory this has super Dynkin labels $[0100010]$; it is
saturated and shortened and can be explicitly written in terms of
$W$ as

 \be
 K_{AB,A'B'}=\del_{(AA'} W \del_{B)B'} W -
 {1\over6}\del_{(AA'}\del_{B)B'} W^2
 \la{32}
 \ee

In the interacting theory it becomes an unconstrained scalar
superfield on Minkowski superspace and acquires an anomalous
dimension \cite{bkrs1,bkrs2}; in analytic superspace it can be
represented as a quasi-tensor superfield (see below) with super
Dynkin labels $[0(1+\c)000(1+\c)0]$ where $\c>0$ is the anomalous
dimension.

The above can be generalised straightforwardly to the case of two
CPOs, $A_{p}, A_{q}$. The OPE is

 \be
 A_p(1) A_q(2) \sim \sum C_{\cR} \cR(X_{12})
 (g_{12})^{{1\over2}(p+q-Q)}
 \cO^Q_{\cR\cR'}(2) +\ldots
 \la{32.1}
 \ee

where the notation is the same as above. Again the operators on the
RHS must be in self-conjugate representations. The constraints
imposed by analyticity in $y_{12}$ are found in the same way as for
two $T$s, but in this case this is not the complete set. To obtain
the other conditions it is convenient to take point 2, say, to be
at the origin, so that the OPE is written entirely in terms of $X$,
the coordinate of $A_p$. Now $A_p$ has a finite expansion in terms
of $y$ which has the highest power $(y^2)^p$. Therefore there is a
bound on the highest power of $y$ that can appear on the right-hand
side. When one takes this into account (for both $A_p$ and $A_q$
taken to be at point 1) one obtains the same constraints as from
the three-point analysis. In the case of two $T$s this procedure
doesn't give any new information.

\section{Quasi-tensor superfields}

In this section we shall discuss the notion of quasi-tensor
superfields which are analytic fields which correspond to
representations with non-integral Dynkin labels over the white
nodes. To make the discussion simpler we shall work in $N=2$ in
$(2,1,1)$ analytic superspace. The super Dynkin diagram for this
space is

 \be
 \begin{picture}(192,10)(0,0) \put(0,0){\makebox[0pt][l]{$\bt
 \hspace {2em}
 \ominus \hspace{2em}\times
 \hspace{2em}\ominus \hspace{2em}\bt$}
 \rule[.5ex]{11.7em}{.1ex} }
 \end{picture}
 \la{33}
 \ee

It is a super Grassmannian with local coordinates

 \be
 X^{AA'}=\left( \ba{cc} x^{\a \adt} & \l^{\a} \\
                        \pi^{\adt}  &  y  \ea
 \right).
\la{34}
 \ee

where the $x$s are the spacetime coordinates, $\l,\pi$ are odd
coordinates and $y$ is the standard local coordinate for the
internal space $\bbC P^1$. We shall put $A=(\a,3)$ and
$A'=(\adt,\dot 3)$. The Levi subalgebra for this superspace is
$\gs\gl(2|1)\oplus\gs\gl(2|1)\oplus \bbC$. The aim is to consider
representations of this algebra which can be extended to
non-integral values of the Dynkin labels over the white nodes.

The simplest example which illustrates this is given by
antisymmetric tensors. Consider a tensor $\cO_{ABC_1\ldots C_p}$
which is (generalised) antisymmetric on all $n=p+2$ indices; it is
not difficult to see that such tensors all have the same number of
components for $p=0,1,2,\ldots$. Tensors of this type carry the
antisymmetric representation of $\gs\gl(2|1)$ in the obvious way.
We now define $\cO_{AB}[p]$ to be the object with components

 \bea
 \cO_{33}[p]&=&\cO_{3333\ldots} \\
 \cO_{\a 3}[p]&=& \cO_{\a 333\ldots} \\
 \cO_{\a\b}[p]&=& \cO_{\a\b 33\ldots}
 \la{35}
 \eea

All of the other components of $\cO$ vanish by antisymmetry of the
even indices. The action of $\gs\gl(2|1)$ is easy to find; if we
take $\gs\gl(2|1)$ to act from the left, we find

 \bea
 \d\cO_{33}[p]&=&(p+2)A_3{}^{\a}\cO_{\a 3}[p]+ (p+2)A_3{}^3\cO_{}[p] \\
 \d\cO_{\a 3}[p]&=&\hat A_{\a}{}^{\b}\cO_{\b 3}[p]+
 (p+{3\over2})A_3{}^3\cO_{\a 3}[p]
 +A_{\a}{}^3\cO_{33}[p]+ (p+1)A_3{}^{\b}\cO_{\a\b}[p]\\
 \d\cO_{\a\b}[p]&=&(p+1)A_3{}^3\cO_{\a\b}[p] -2A_{[\a}{}^3\cO_{\b]3}[p]
 \la{36}
 \eea

where $A_A{}^B\in\gs\gl(2|1)$ and $\hat A_{\a}{}^{\b}$ is
traceless. Now this formula makes sense for arbitrary values of $p$
(even complex) and it is straightforward to check that one still
has a representation of the algebra. In the context of the $N=2$
superconformal group, unitarity requires $p$ to be real and $p\geq
-1$. (Note that $p$'s being less than zero is not a problem for
$\cO_{\a\b}[p]$; we could, if we wished, write this as
$\cO_{\a\b}[p]=\e_{\a\b}\cO[p]$ and eliminate the antisymmetrised
$\a\b$ index pair altogether.) There is just one special value of
$p$, namely where $p=-1$. For this value one sees that the
representation becomes reducible, although not completely
reducible. The components $(\cO_{33}[-1], \cO_{\a 3}[-1])$
transform under a subrepresentation, while the quotient
representation in this case is just a singlet. It is just this
phenomenon which is happening in the case of the $N=4$ Konishi
operator. In the interacting quantum theory the odd Dynkin labels
are $2+p$ where $p=\c-1$; in the free theory $\c=0$ which implies
$p=-1$ at which point the subrepresentation is the free Konishi
multiplet.

Returning to $N=2$ we can now consider an operator in the theory
which transforms under the same antisymmetric representation of
both $\gs\gl(2|1)$'s. In the OPE of two $N=2$ CPOs such an operator
could appear on the RHS if we could define $R(X_{12})$ for
non-integral $p$. This is not difficult to do as we now explain.
The antisymmetrised product of $n$ $X$'s, $X^{[A A'}X^{BB'}\ldots
X^{C_p]C'_p}$,  again has the property that most of the components
vanish. We can therefore arrange these non-vanishing components in
a $(4|4)\xz (4|4)$ matrix which we denote by $(X^n)^{AB,A'B'}$.
This is antisymmetric on both pairs of indices. The components of
this matrix are

 \be
 \left(\ba{lll}
 (\hy+\p x^{-1}\l)^n
  & \p^{\adt} (\hy+\p x^{-1}\l)^{n-1}
   & \p^{\adt}\p^{\bdt}\hy^{n-2} \\
  &&\\
 \l^{\a}(\hy+\p x^{-1}\l)^{n-1}
 & {\hy^{n-2}\over n}
 \left(x^{\a\adt}(\hy +(n-1)\p x^{-1}\l)
 +(n-1)\l^{\a}\p^{\adt}\right)
 &{2\over n}
 x^{\a[\adt}\p^{\bdt]}\hy^{n-2} \\
 &&\\
 \l^{\a}\l^{\b}\hy^{n-2}& {2\over n}x^{[\a \adt}\l^{\b]}\hy^{n-2}&{2\over
 n(n-1)}x^{[\a\adt} x^{\b]\bdt}\hy^{n-2}
 \ea\right)
 \la{37}
 \ee

where the index arrangement is

 \be
 [AB][A'B']=\left(\ba{rrr}
 33,\3dt\3dt & 33,\adt\3dt & 33,\adt\bdt \\
 \a 3,\3dt\3dt & \a 3,\adt\3dt & \a 3, \adt\bdt \\
 \a\b,\3dt\3dt & \a\b,\adt\3dt & \a\b,\adt\bdt
 \ea
 \right)
 \la{38}
 \ee

We observe that there is a common factor of $\hy^{n-2}=\hy^p$. When
this is factored out the remaining matrix is analytic in $\hy$ and
depends on $p$ only through numerical coefficients. The matrix can
now be defined for arbitrary values of $p$ and can appear on the
RHS of the OPE of two CPOs contracted with the operator discussed
above. The key point is that there will be a factor of $\hy^{-p}$
coming from the propagator factor in the OPE so that the
non-analyticity in $\hy$ will cancel out. On the other hand, there
will be non-integral powers of $x^2$ so that we can have anomalous
dimensions in a way which is perfectly compatible with analyticity.

\section{$U(1)_Y$}

In \cite{i} it was argued that there might be a bonus $U(1)$
symmetry of some $N=4$ SYM correlation functions whose origin could
be traced to IIB supergravity. This idea was extended in \cite{is} where it was conjectured that the OPE ivolving at least two short operators might be $U(1)_Y$ invariant and that this would imply the non-renormalisation theorems for two- and three-point functions of CPOs. We
recall that these non-renormalisation theorems had been found on
the AdS side \cite{ads} and checked in certain, mainly
perturbative, field theory calculations \cite{nonren,hsw,bkrs2}.
It was further noted in \cite{i} that the analytic superconformal invariants listed in
\cite{hw4} were invariant under this additional symmetry. It was
realised in \cite{ehw} that, using the reduction formula introduced
in the superconformal context in \cite{i}, one could construct a
proof of the non-renormalisation theorem for two- and three-point
functions of CPOs which can be interpreted in terms of $U(1)_Y$.  As
shown in \cite{ehw,hssw,hw} there are additional, nilpotent
invariants which do not need to be invariant under this symmetry
which were omitted from the invariants listed in \cite{hw4}.
However, these invariants occur at five or more points, so that the
$n$-point functions of CPOs for $n\leq 4$ are indeed invariant
under this additional symmetry. In this section we extend this
discussion to arbitrary protected operators in $N=4$ SYM. The
conclusions are the same as for CPOs: the two-, three- and
four-point functions are $U(1)_Y$ invariant and the two- and
three-point functions should be non-renormalised.

To see how this symmetry arises, we note that $(4,2,2)$ analytic
superspace can be regarded as a coset space of $GL(4|4)$. However,
when one works out the action of the group on the coordinates $X$
one finds that (infinitesimal) transformations proportional to the
unit matrix do not act, so that one naturally obtains an action of
$PGL(4|4)$. Explicitly, if the constraint \eq{12} is dropped, we
extend the transformation group in precisely this way. If we denote
the diagonal elements of the super matrices $A$ and $D$ by

 \be
 A\sim {1\over 2}\left(\ba{cc}
 a_o I_2 & 0\\
 0& a_1 I_2\ea\right)
 \la{39}
 \ee

and similarly for $D$ we find the transformations

 \bea
 \d x&=& s x \\
 \d\l&=& {1\over 2}(s + s' + t)\l\\
 \d\p&=& {1\over 2}(s + s' -t)\l\\
 \d y&=& s' y
 \la{40}
 \eea

where

 \bea
 s&=& {1\over2} (a_o + d_o)\\
 s'&=& {1\over2} (a_1 + d_1)\\
 t&=&{1\over2}(a_o-a_1-(d_o-d_1))=-{1\over2}\str(\d g)
 \la{41}
 \eea

The parameters $s$ and $s'$ correspond to dilations and internal
dilations respectively, and we can identify $t$ as the $U(1)_Y$
parameter which acts only on the odd variables. If we enlarge the
group in this way we find that

 \be
 \D(s,s',t)=s-s' + t
 \la{42}
 \ee

The basic propagator $g_{12}$, which can be interpreted as the
two-point function of two $W$s in the free theory, will be
invariant under the enlarged symmetry if we assign $U(1)_Y$ charge
$-1$ to $W$. This simply has the effect of cancelling the $t$ term
from $\D$, so that invariance follows because $g_{12}$ depends on
the odd variables only as a product $\l\p$.

Now consider an arbitrary protected analytic tensor operator
$\cO^Q_{\cR\cR'}$. If we assign it $U(1)_Y$ charge $-Q$ then the
parameter of this transformation will only appear via the matrices
$A(X)$ and $D(X)$ in \eq{7}. Therefore the Ward identities for two-
and three-point functions of such operators will still be satisfied
for this larger symmetry group because verifying that this is the
case is essentially the same as for $PSL(4|4)$.

The reduction formula states that the derivative of an $n$-point
correlation function with respect to the complex coupling constant
$\t$ is given by an $(n+1)$-point function which includes an
integrated insertion of the on-shell action \cite{i}. In analytic
superspace this formula can be written as \cite{ehw}

 \be
 {\del\over\del\t}<\cO_1\ldots\cO_n>\sim \int\, d\m_o\,
 <T_o\cO_1\ldots\cO_n>
 \la{43}
 \ee

where $d\m=d^4x\,d^4 y\, d^4\l$. This measure is easier to
interpret in harmonic superspace, but for the following argument
the key point is that the integrand will need a factor of $\l_o^4$
in order to obtain a non-zero integral. $T_o$ is the supercurrent
inserted at point $0$.\footnote{The fact that the on-shell action
can be written as integral over a subspace of $N=4$ superspace was
noted some time ago \cite{hst}.} If we apply this formula to a
two-or three-point function of protected analytic operators the
integrand on the RHS will involve a three- or four-point function
of protected operators with the same tensorial structure as the
left-hand side since $T_o$ has no superindices. Now invariance of
the left-hand side for a given component will involve given powers
of the $\l$s and $\p$s as they are the only coordinates which carry
$U(1)_Y$ charge. These powers must be the same on the right-hand
side because the tensorial structure is the same. But to obtain a
non-zero integral we need an extra four powers of $\l_o$, and this
is clearly not possible. Thus the integrals must be zero (there may
be a sum of terms on the right) and we conclude that the two- or
three-point function on the left-hand-side must be
non-renormalised.

The above argument is predicated on the fact that the integrands
are $U(1)_Y$ invariant. This is true for the three-point functions
by construction, but we need to show that it is also true for
four-point functions of protected operators. For $n$-point
functions of tensor operators we can proceed as for two- or
three-points, namely we start by translating all the points to
point 1, say, by means of $X^{-1}_{1i}, i=2,\ldots n$. We can then
take care of the dilation weights by multiplying by appropriate
factors of the propagators. We shall then obtain solutions to the
Ward identities by multiplying by tensors formed from monomials of
the coordinate functions $X_{12j}:=X_{12} X^{1}_{2j} X_{j1},\,
j=2,\ldots n$ and their inverses. There may be many independent
solutions of this type and we can multiply each one of them by a
function of the invariants. By construction, these tensorial
functions will be $U(1)_Y$ invariant, so to prove the
non-renormalisation theorem for three points we only need to
observe that the four-point invariants are themselves invariant
under $U(1)_Y$. The argument breaks down for four-point functions
because there are nilpotent five-point invariants which are not
$U(1)_Y$ invariant, as we remarked above.

\section{Extremal correlators}

We recall that extremal correlators are by definition correlation functions 
of CPOs such that the charge (central super Dynkin label) at one point equals the sum of charges at all of the other points. These were shown to have a free-field functional form and to be non-renormalised on the AdS side in \cite{dfmmr}, and this was checked on the field theory side in perturbation theory \cite{bk} and non-perturbatively \cite{ehssw}. There are also next-to-extremal correlators which exhibit similar behaviour \cite{ehssw,ep,afro,ehsw}. In \cite{es,efs} this behaviour of extremals has been interpreted from the point of view of OPEs.

To be explicit consider a simple example, $<A_6(1) T(2) T(3) T(4)>$. If one carries out OPE expansions in $(12)$ and $(34)$, then, using the restrictions on the OPES which are due to analyticity, it is easy to show that only short protected operators can contribute in the final two-point function. It follows from this that the functional form of the extremal correlator is a product of free propagators multiplied by a coefficient which could, in principle, depend on the coupling. However, the same argument shows that this depedence must be trivial, because the OPE coeficients  and the coefficients of the two-point functions of the exchanged operators which contribute when one carries out the double OPE expansion are determined by two- or three-point functions of protected operators.

We shall now argue that this picture can be generalised to protected operators other than CPOs. To be explicit we shall consider an example of a four-point correlation function consisting of three $T$s and a series C operator with Dynkin labels $[1p1]$. We recall that such an operator is represented on analytic superspace by a covector operator $\cO^{p+2}_{A'A}$. The claim is that such a correlation function is extremal if $p=4$, i.e. $Q=6$. In this case we can write

\be
<\cO^{6}_{A'A}(1) T(2) T(3) T(4)>=(f_{123}(X^{-1}_{123})_{A'A} +f_{124}(X^{-1}_{124})_{A'A})\xz (g_{12} g_{13} g_{14})^2
\la{43.1}
\ee

where $f_{123}$ and $f_{124}$ are two arbitrary functions of the invariants. If we again carry out the double OPE expansion on this correlator we find as before that the only operators which can contribute in the intermediate channel are protected. 

To show this we can either analyse the $T\cO^6_{A'A}$ OPE or three-point functions of the form  \linebreak $<\cO^Q_{RR'}\cO^6_{AA'}T>$. We choose to do the latter. There are two possible solutions for this three-point function:

\bea
(i)\ <\cO^Q_{\unB'\unB}(1)\cO^{6}_{A'A}(2) T(3)>&=& 
P\xz\cR\left( (X^{-1}_{12})_{B_1'A} (X^{-1}_{12})_{A'B_1}\cR_1(X^{-1}_{123})_{\unB'_2\unB_2}\right)\\
&&\nn\\
(ii)\ <\cO^Q_{\unB'\unB}(1)\cO^{6}_{A'A}(2) T(3)>&=& P\xz
(X^{-1}_{231})_{A'A}\cR(X^{-1}_{123})
\la{43.2}
\eea

where in both cases the propagator factor $P$ is given by

\be
P=(g_{12})^{{Q\over2}+2}(g_{13})^{{Q\over2}-2} (g_{23})^{4-{Q\over2}}
\la{43.3}
\ee

The notation in \eq{43.2} is as follows: $\unB(\unB')$ stands for all the unprimed (primed) indices in the representation $\cR(\cR')$ while $\unB_2$ stands for one fewer index in the representation $\cR_1$ which is self-conjugate; the operation $\cR$ on the RHS of (i) then puts all of the $\unB(\unB')$ indices within the bracket into the representations $\cR(\cR')$.

We can now show that analyticity implies there can be no unprotected operators ($b\geq2$) which can contribute to the three-point function and which can appear in the OPE of two $T$s. In fact, one can show that there are no solutions of type (ii) with $b\geq2$ which are compatible with analyticity. For solutions of type (i) with $b\geq2$ one can show that either $d=2$ or $d=4$ whereas the only unprotected operators in $TT$ have $d=0$. The operators which can contribute to both the three-point function and the $TT$ OPE can also be found. The only possible contribution is type (i) with super Dynkin labels $[0004000]$, i.e. a short series C operator with $SU(4)$ labels $[040]$.

A similar calculation shows that no unprotected operators are exchanged in the next-to-extremal case, $<\cO^4_{A'A}TTT>$.

The example given here can be generalised to the case of a $[1p1]$ operator with three other chiral primaries whose charges sum to $Q=p+2$. Presumably there are many more extremal and next-to-extremal correlators.

\section{Conclusions}

In this paper we have exploited the fact that all of the unitary
representations of $PSU(4|4)$ that arise in $N=4$ SYM can be
realised as analytic superfields on $(4,2,2)$ superspace. Broadly
speaking they divide into two categories, the protected operators,
which are represented by analytic tensor superfields and which have
integral super Dynkin labels, and the unprotected operators, which
can be realised by what we have called quasi-tensor superfields and
which have non-integral labels for at least one of the white (odd)
nodes in the particular super Dynkin basis we have been using. We
have given explicit formulae for the two- and three-point functions
valid, in principle, for any protected operators and have derived
how these operators appear in the OPEs of any two chiral primary
operators. These OPE formulae can clearly be extended to more
complicated protected operators without too much difficulty. We
have also seen that $n$-point  correlation functions of protected
operators for $n\leq 4$ are invariant under an additional $U(1)_Y$
symmetry and have used this fact to argue that the two- and
three-point functions should be non-renormalised.

In section 6 we showed how unprotected operators can be
accommodated in analytic superspace in an $N=2$ example. This
result shows how such operators can occur in the OPE of two
protected operators and leads us to believe that the two-and
three-point correlators and OPE formulae given in the paper are
probably also valid for the unprotected operators provided that
factors such as $\cR(X_{12})$ are interpreted appropriately. It would be interesting to use this formalism to try to prove the conjecture of \cite{is} concerning the $U(1)_Y$ behaviour of OPES of different types of operator, but so far we have not done this.

Three-point functions with at least one unprotected operator cannot
be non-renormalised, and it is not so clear that they are invariant
under $U(1)_Y$ (although see comments in \cite{is}). 
However, it might be possible to  employ the
reduction formula to obtain some information about the
$\t$-dependence of such correlators - one might imagine being able
to relate the $\t$-dependence of the coefficient of such a function
to the anomalous dimension of the operator, for example.

Finally, the methods advocated in this paper may have further
applications. For example, we are now in a position to analyse
four- and higher point functions directly in analytic superspace
using the OPE. Another application would be to superconformal field
theories in other spacetime dimensions, particularly $D=6,(2,0)$
and $D=3, N=8$ SCFT. In a recent paper \cite{efs} it has been shown
that there are also protected series A operators in $D=6, (2,0)$
and this has been used to give a proof of the triviality of certain
extremal correlators studied on the AdS side in \cite{dp} and in
analytic superspace in \cite{h}.

{\bf Acknowledgement}

This research was supported in part by PPARC SPG grant 613.

\end{document}